\documentclass[12pt]{article}

\newcommand{\be}{\begin{equation}}
\newcommand{\ee}{\end{equation}}
\newcommand{\bi}[1]{\vspace{-2mm} \bibitem{#1}}

\usepackage{epsfig,amsmath,amssymb,graphics,graphicx}

\begin{document}

\begin{center}
{\it Modern Physics Letters B 21 (2006) 237-248}
\end{center}

\begin{center}
{\Large \bf Liouville and Bogoliubov Equations \\
with Fractional Derivatives}
\vskip 5 mm

{\large \bf Vasily E. Tarasov} \\

\vskip 3mm

{\it Skobeltsyn Institute of Nuclear Physics, \\
Moscow State University, Moscow 119991, Russia } \\
{E-mail: tarasov@theory.sinp.msu.ru}
\end{center}


\begin{abstract}
The Liouville equation, first Bogoliubov hierarchy 
and Vlasov equations with derivatives of non-integer order
are derived.
Liouville equation with fractional derivatives 
is obtained from the conservation of probability  
in a fractional volume element.
This equation is used to obtain Bogoliubov hierarchy and 
fractional kinetic equations with fractional derivatives.
Statistical mechanics of fractional generalization of the 
Hamiltonian systems is discussed.
Fractional kinetic equation for the system of 
charged particles are considered. 
\end{abstract}


\section{Introduction}

Fractional equations \cite{Podlubny} are equations that contain 
derivatives of non-integer order \cite{OS,SKM}. 
The theory of derivatives of non-integer order goes back 
to Leibniz, Liouville, Riemann, and Letnikov \cite{SKM}. 
Derivatives and integrals of fractional order have found many
applications in recent studies in mechanics and physics.
In a short period of time the list of applications have become long.
For example, it includes 
chaotic dynamics \cite{Zaslavsky1,Zaslavsky2},
mechanics of fractal media \cite{Mainardi,Hilfer,Media},
physical kinetics \cite{Zaslavsky1,Zaslavsky7,SZ,ZE,Nigmat},
plasma physics \cite{CLZ,Mil2,Plasma2005}, 
astrophysics \cite{CMDA},
long-range dissipation \cite{GM,M,TZ2}, 
mechanics of non-Hamiltonian systems \cite{nonHam,FracHam},
theory of long-range interaction \cite{Lask,TZ3,KZT}, 
and many others physical topics.

In this paper, we derive Liouville equation with fractional 
derivatives with respect to coordinates and momenta. 
To derive fractional Liouville equation, we consider the conservation
of probability to find a system in the fractional differential volume element.
Using the fractional Liouville equation, we derive
the fractional generalization of the Bogoliubov hierarchy equations.
These equations can be used to derive 
fractional kinetic equations \cite{Zaslavsky1,Zaslavsky7,SZ,Nigmat}.
A linear fractional kinetic equation for the system
of charged particles is suggested. 

In Sec. 2, we derive the Liouville equation with 
fractional derivatives from the conservation of probability 
to find a system in the fractional volume element of the phase space. 
In Sec. 3, we obtain the first Bogoliubov hierarchy equation 
with fractional derivatives in the phase space from the fractional 
Liouville equation.
In Sec. 4, the Vlasov equation 
with fractional derivatives in phase space is considered.
In Sec. 5, a linear fractional kinetic equation for the system
of charged particles is suggested. 
Finally, a short conclusion is given in Sec. 6.

\section{Liouville equation with fractional derivatives}

A basic principle of statistical mechanics 
is the conservation of probability.
The Liouville equation is an expression of this basic 
principle in a convenient form for the analysis.
In this section, we derive the Liouville equation with 
fractional derivatives from the conservation of probability  
in a fractional volume element.

For the phase space $\mathbb{R}^{2n}$ with coordinates 
$(x^1,...,x^{2n})=(q_1,...,q_n,p_1,...,p_n)$, we
consider a fractional differential volume element 
\be \label{1}
d^{\alpha} V = d^{\alpha} x_1 ... d^{\alpha} x_{2n} .
\ee
Here, $d^{\alpha}$ is a fractional differential \cite{FDF1}. 
For the function $f(x)$,
\be \label{fd}
d^{\alpha} f(x)=\sum^{2n}_{k=1} D^{\alpha}_{x_k} f(x) (d x_k)^{\alpha},
\ee
where $D^{\alpha}_{x_k}$ is a fractional derivative \cite{SKM}
of order $\alpha$ with respect to $x_k$. 
The Caputo derivative \cite{Caputo2,Mainardi} is defined by 
\be
D^{\alpha}_xf(x)=\frac{1}{\Gamma(m-\alpha)} 
\int^x_0 \frac{f^{(m)}(z)}{(x-z)^{\alpha+1-m}} d z ,
\ee
where $m-1 < \alpha < m$, \ $f^{(m)}(\tau)=d^m f(\tau)/d\tau^m$, and
$\Gamma(z)$ is the Euler gamma-function.
For Caputo and Riesz Eq. \cite{SKM} 
fractional derivatives, we have $D^{\alpha}_{x_k} 1=0$, and 
$D_{x_k}x_l=0$ $(k\not=l)$. 
Using (\ref{fd}), we obtain
\be
d^{\alpha} x_k=D^{\alpha}_{x_k} x_k  (d x_k)^{\alpha}, 
\quad (\alpha >0).
\ee
Then
\be \label{Eq5*}
(d x_k)^{\alpha} = \left( D^{\alpha}_{x_k} x_k \right)^{-1} d^{\alpha} x_k .
\ee
For Caputo derivatives, 
\be D^{\alpha}_{x_k} x^{\beta}_k=
\frac{\Gamma(\beta+1)}{\Gamma(\beta+1-\alpha)} x^{\beta-\alpha}_k  ,
\quad ( \beta>\alpha ) , \ee
we get
\be
(d x_k)^{\alpha} = \left( D^{\alpha}_{x_k} x_k \right)^{-1} d^{\alpha} x_k 
=\Gamma(2-\alpha) x^{\alpha-1}_k d^{\alpha} x_k .
\ee

The conservation of probability to find a many-particle system in 
the phase-space volume element $d^{\alpha}V$ 
may be expressed as
\be 
-d V \frac{\partial \rho(t,x)}{\partial t}=
d [\rho(t,x) \; ({\bf u},d{\bf S}) ] 
\ee
for the usual volume element ($\alpha=1$), and
\be \label{duS}
-d^{\alpha} V \frac{\partial \rho(t,x)}{\partial t}=
d^{\alpha} [\rho(t,x) ({\bf u},d^{\alpha}{\bf S}) ] 
\ee
for non-integer $\alpha$.
Here, $\rho=\rho(t,x)$ is the density of probability
to find a many-particle system in the phase-space volume element; 
${\bf u}={\bf u}(t,x)$ is the velocity vector field,
$d^{\alpha} {\bf S}$ is a surface element, and
the brackets $( \ , \ )$ is a scalar product of vectors:
\be \label{uds}
{\bf u}=\sum^{2n}_{k=1} u_{k} {\bf e}_k , \quad
d^{\alpha} {\bf S}= \sum^{2n}_{k=1} d^{\alpha} S_k {\bf e}_k ,
\quad
({\bf u},d^{\alpha}{\bf S}) =\sum^{2n}_{k=1} u_k d^{\alpha} S_k ,
\ee
where ${\bf e}_k$ are the basic vectors of Cartesian coordinate system, and
\be \label{daS}
d^{\alpha}S_k =d^{\alpha} x_1 ...d^{\alpha} x_{k-1} d^{\alpha} x_{k+1}  ...
d^{\alpha} x_{2n} ,
\ee
The functions  $u_k=u_k(t,x)$ define $x_k$ components of 
the velocity vector field ${\bf u}(t,x)$, 
the rate at which probability density is transported
through the area element $d^{\alpha}S_k$.
In the usual case ($\alpha=1$), the outflow of probability 
in the $x_k$ direction is defined by
\be \label{dru1}
d [\rho u_k ] d S_k =D_{x_k} [\rho u_k ] dx_k d S_k 
=D_{x_k} [\rho u_k ]  d V .
\ee
The fractional generalization of Eq. (\ref{dru1}) is
\[
d^{\alpha} [\rho u_k ] d^{\alpha} S_k 
=D^{\alpha}_{x_k} [\rho u_k ] (dx)^{\alpha} d^{\alpha} S_k= \]
\be \label{dru1a}
=\left( D^{\alpha}_{x_k} x_k \right)^{-1} D^{\alpha}_{x_k} [\rho u_k ] 
d^{\alpha} x_k d^{\alpha} S_k =
\left( D^{\alpha}_{x_k} x_k \right)^{-1}
D^{\alpha}_{x_k} [\rho u_k ]  d^{\alpha} V .
\ee
Here, we use Eqs. (\ref{daS}), (\ref{1}) and (\ref{Eq5*}).
Substitution of Eq. (\ref{dru1a}) into Eq. (\ref{duS}) gives
\be
-d^{\alpha} V \frac{\partial \rho}{\partial t}=
d^{\alpha} V \sum^{2n}_{k=1}
\left( D^{\alpha}_{x_k} x_k \right)^{-1} D^{\alpha}_{x_k} [\rho u_k ] . 
\ee

As a result, we obtain 
\be \label{cont1}
\frac{\partial \rho}{\partial t}=
-\sum^{2n}_{k=1} \left( D^{\alpha}_{x_k} x_k \right)^{-1}
D^{\alpha}_{x_k} [\rho u_k ] .
\ee
For Caputo derivatives
\be \label{cont1a}
\frac{\partial \rho}{\partial t}=
-\Gamma(2-\alpha) \sum^{2n}_{k=1} x^{\alpha -1}_k 
D^{\alpha}_{x_k} [\rho u_k ] .
\ee
Equation (\ref{cont1}) is the Liouville equation
that contains the derivatives of fractional order $\alpha$.
Fractional Liouville equation (\ref{cont1}) 
describes the probability conservation to find a system in
the fractional volume element (\ref{1}) of the phase space.

For the coordinates 
$(x^1,...,x^{2n})=(q_1,...,q_n,p_1,...p_n)$,
Eq. (\ref{cont1}) is
\be \label{Liouv1}
\frac{\partial \rho}{\partial t}
+\sum^{n}_{k=1}\left( \left( D^{\alpha}_{q_k} q_k \right)^{-1} 
D^{\alpha}_{q_k} [\rho V_k ] \right)
+\sum^{n}_{k=1}\left( \left( D^{\alpha}_{p_k} p_k \right)^{-1}
D^{\alpha}_{p_k} [\rho F_k ] \right)=0,
\ee
where $V_k=u_k$, and $F_k=u_{k+n}$ ($k=1,...,n$).
The functions $V_k=V_k(t,q,p)$ are the components of velocity field,  
and $F_k=F_k(t,q,p)$ are the components of force field.
In general,  
\be
D^{\alpha}_{p_k}[\rho F_k ] \not=
\rho D^{\alpha}_{p_k} F_k +F_k D^{\alpha}_{p_k}\rho.
\ee
Suppose that $F_k$ does not depend on $p_k$, and
the $k$th component $V_k$ of the velocity field does not 
depend on $k$th component $q_k$ of coordinates.
In this case, Eq. (\ref{Liouv1}) gives
\be \label{Liouv2}
\frac{\partial \rho}{\partial t}
+\sum^{n}_{k=1} V_k \left( D^{\alpha}_{q_k} q_k \right)^{-1}
D^{\alpha}_{q_k} \rho 
+\sum^{n}_{k=1} F_k \left( D^{\alpha}_{p_k} p_k \right)^{-1}
D^{\alpha}_{p_k} \rho =0.
\ee
If we consider the fractional generalization of 
Hamiltonian system \cite{FracHam}, 
then $V_k$ and $F_k$ can be represented as fractional
derivatives of some function $H(q,p)$:
\be \label{VkFk}
V_k=D^{\alpha}_{p_k} H(q,p), \quad
F_k=-D^{\alpha}_{q_k} H(q,p). 
\ee
For $\alpha=1$, we have the usual Hamiltonian system. 

The classical system that is defined by the equations
\be \label{HamDef}
\frac{dq_k}{dt}=V_k(t,q,p), \quad \frac{dq_k}{dt}=F_k(t,q,p) 
\ee
is called a Hamiltonian system, if the right-hand sides 
of equations (\ref{HamDef}) satisfy the Helmholtz conditions  
\be \label{HC1} \frac{\partial V_{i}}{\partial p_j}-
\frac{\partial V_{j}}{\partial p_i}= 0, \quad
\frac{\partial V_{j}}{\partial q_i}+
\frac{\partial F_{i}}{\partial p_j}=0, \quad
\frac{\partial F_{i}}{\partial q_j}-
\frac{\partial F_{j}}{\partial q_i}= 0 . \ee
In this case,  Eq. (\ref{HamDef}) can be presented in the form
\be \frac{dq_k}{dt}=\frac{\partial H}{\partial p_k}, 
\quad \frac{dp_k}{dt}=-\frac{\partial H}{\partial q_k}  \ee
that is uniquely defined by the Hamiltonian $H$.
The fractional Hamiltonian systems are defined in Ref. \cite{nonHam}.

For Eq. (\ref{VkFk}), equation (\ref{Liouv2}) is
\be \label{Liouv3}
\frac{\partial \rho}{\partial t}
+\sum^{n}_{k=1} \left( D^{\alpha}_{q_k} q_k \right)^{-1}
D^{\alpha}_{p_k} H  D^{\alpha}_{q_k} \rho 
-\sum^{n}_{k=1} \left( D^{\alpha}_{p_k} p_k \right)^{-1}
D^{\alpha}_{q_k} H D^{\alpha}_{p_k} \rho =0.
\ee
We can define 
\be \label{br}
\{ A,B \}_{\alpha}=\sum^n_{k=1} 
\left( \left( D^{\alpha}_{q_k} q_k \right)^{-1} D^{\alpha}_{q_k} A 
D^{\alpha}_{p_k} B
- \left( D^{\alpha}_{p_k} p_k \right)^{-1}
D^{\alpha}_{q_k} B D^{\alpha}_{p_k} A  \right) .
\ee
For $\alpha=1$, Eq. (\ref{br}) defines Poisson brackets.
Note that the brackets (\ref{br}) satisfy the relations
\[ \{A,B\}_{\alpha}=- \{B,A\}_{\alpha}, \quad \{1,A\}_{\alpha}=0 .\]
In general, the Jacoby identity  cannot be satisfied.
The property $\{1,A\}_{\alpha}=0$ is satisfied only for
Caputo and Riesz fractional derivatives ($D^{\alpha}_x1=0$).
For the Riemann-Liouville derivative, $D^{\alpha}_x1\not=0$.
Using Eq. (\ref{br}), we get Eq. (\ref{Liouv3}) in the form
\be
\label{Liouv5}
\frac{\partial \rho}{\partial t}+\{\rho,H\}_{\alpha}=0.
\ee
As a result, we have the Liouville equation 
for the fractional generalization of Hamiltonian systems \cite{FracHam}
that are defined by Eq. (\ref{VkFk}).
For $\alpha=1$, Eq. (\ref{Liouv5}) is the usual 
Liouville equation for Hamiltonian systems.

\section{Bogoliubov equation with fractional derivatives}

Let us consider a classical system with a fixed number $N$ of
identical particles. 
Suppose the $k$th particle is described by the 
generalized coordinates $q_{ks}$ and generalized
momenta $p_{ks}$, where $s=1,...,m$.
We use the notations
${\bf q}_k=(q_{k1},...,q_{km})$ and
${\bf p}_k=(p_{k1},...,p_{km})$.
The state of this system is described by the N-particle density 
of probability $\rho_{N}$ in the $2mN$-dimensional phase space:
\[ \rho_{N}({\bf q},{\bf p},t)=
\rho({\bf q}_{1},{\bf p}_{1},...,{\bf q}_{N},{\bf p}_{N},t). \]
The fractional Liouville equation is
\be \label{r2}
\frac{\partial \rho_{N}}{\partial t}=-
\sum^{N}_{k=1} \Bigl(
{\bf D}^{\alpha}_{\bf q_k} ({\bf V}_k \rho_{N})+
{\bf D}^{\alpha}_{\bf p_k} ({\bf F}_k \rho_{N}) \Bigr) , \ee
where 
\be \label{bfD}
{\bf D}^{\alpha}_{\bf q_k} {\bf V}_k=\left( D^{\alpha}_{{\bf q}_k} 
{\bf q}_k \right)^{-1} D^{\alpha}_{\bf q_k}{\bf V}_k=
\sum^m_{s=1} \left( D^{\alpha}_{q_{ks}} q_{ks} \right)^{-1} 
D^{\alpha}_{q_{ks}} V_{ks},
\ee
\be
{\bf D}^{\alpha}_{\bf p_k}{\bf F}_k
=\left( D^{\alpha}_{{\bf p}_k} {\bf p}_k \right)^{-1} 
D^{\alpha}_{\bf p_k} {\bf F}_k
=\sum^m_{s=1} \left( D^{\alpha}_{q_{ks}} p_{ks} \right)^{-1} 
D^{\alpha}_{p_{ks}} F_{ks} .
\ee
The one-particle reduced density of probability
$\rho_1$ can be defined by 
\be \label{r1}  \rho_{1}({\bf q},{\bf p},t)=
\rho({\bf q}_{1},{\bf p}_{1},t)=
\hat I [2,...,N] \rho_{N}({\bf q},{\bf p},t), \ee
where $\hat I[2,...,N]$ is an integration with respect 
to variables ${\bf q}_2$, ..., ${\bf q}_N$, ${\bf p}_2$, ..., ${\bf p}_N$.
Obviously, that one-particle density of probability satisfies
the normalization condition 
\be \label{r3} 
\hat I[1]  \rho_{1}({\bf q},{\bf p},t)=1 . \ee

The Bogoliubov hierarchy equations \cite{Bog3,Bog3b,Gur,Petrina}
describe the evolution of the reduced density of probability.
They can be derived from the Liouville equation.
Let us derive the first Bogoliubov equation with fractional 
derivatives from the fractional Liouville equation (\ref{r2}).
Differentiation of Eq. (\ref{r1}) with respect to time gives
\[ \frac{\partial  \rho_{1}}{\partial t}=
 \frac{\partial}{\partial t} \hat I[2,...,N]  \rho_{N}=
\hat I[2,...,N] \frac{\partial  \rho_{N}}{\partial t} . \]
Using Eq. (\ref{r2}), we get
\be \label{r1i}
\frac{\partial  \rho_{1}}{\partial t}=
- \hat I[2,...,N] \sum^{N}_{k=1} \Bigl(
{\bf D}^{\alpha}_{\bf q_k} ({\bf V}_{k} \rho_{N})+
{\bf D}^{\alpha}_{\bf p_k} ({\bf F}_{k} \rho_{N})\Bigr) . 
\ee

Let us consider the integration over
${\bf q}_{k}$ and ${\bf p}_{k}$ for k-particle term of Eq. (\ref{r1i}). 
Since the coordinates and momenta are independent variables, 
we derive
\be \label{lim} \hat I[{\bf q}_k]
{\bf D}^{\alpha}_{\bf q_k} ({\bf V}_k  \rho_{N}) =
\hat I[{\bf q}_k] \left( D^{\alpha}_{{\bf q}_k} {\bf q}_k \right)^{-1}
D^{\alpha}_{\bf q_k} ({\bf V}_k  \rho_{N}) \sim
\Bigl({\bf V}_k  \rho_{N} \Bigr)^{+\infty}_{-\infty}=0 . \ee
For example, the Caputo derivatives give
\[ \hat I[{\bf q}_k]
{\bf D}^{\alpha}_{\bf q_k} ({\bf V}_k  \rho_{N}) =
\Gamma(2-\alpha)\hat I[{\bf q}_k] {\bf q}^{\alpha-1}_k
D^{\alpha}_{\bf q_k} ({\bf V}_k  \rho_{N})= \]
\be \label{lim2}
=\Gamma(\alpha)\Gamma(2-\alpha)\hat I^{\alpha}[{\bf q}_k] 
D^{\alpha}_{\bf q_k} ({\bf V}_k  \rho_{N})=
\Gamma(\alpha)\Gamma(2-\alpha)
\Bigl({\bf V}_k  \rho_{N} \Bigr)^{+\infty}_{-\infty}=0 , \ee
where $\hat I^{\alpha}[{\bf q}_k]$ is a fractional integration
with respect to variables ${\bf q}_k$.
In Eq. (\ref{lim}), we use the fact 
that the density of probability $\rho_{N}$ in the limit
${\bf q}_k \rightarrow \pm \infty$ is equal to zero.
It follows from the normalization condition.  
If the limit is  not equal to zero, then
the integration over the phase space is equal to infinity.
Similarly, we obtain
\[ \hat I[{\bf p}_{k}] 
{\bf D}^{\alpha}_{\bf p_k} \Bigl({\bf F}_{k}  \rho_{N} \Bigr) \sim
\Bigl({\bf F}_{k}  \rho_{N} \Bigr)^{+\infty}_{-\infty}=0  . \]
Then all terms in Eq. (\ref{r1i}) with $k=2,...,N$
are equal to zero. We have only the term with $k=1$.
Therefore Eq. (\ref{r1i}) has the form
\be \label{r1i2} \frac{\partial  \rho_{1}}{\partial t}=
- \hat I[2,...,N]\Bigl(
{\bf D}^{\alpha}_{\bf q_1} ({\bf V}_1  \rho_{N})
+{\bf D}^{\alpha}_{\bf p_1}({\bf F}_{1}  \rho_{N})
\Bigr) . \ee

Since the variable ${\bf q}_{1}$ is an independent of
${\bf q}_{2},...,{\bf q}_{N}$ and  ${\bf p}_{2},...,{\bf p}_{N}$,
the first term in Eq. (\ref{r1i2}) can be written as
\[ \hat I[2,...,N]
{\bf D}^{\alpha}_{\bf q_k} ({\bf V}_1  \rho_{N}) 
= {\bf D}^{\alpha}_{\bf q_1}
{\bf V}_1 \hat I[2,...,N]  \rho_{N} =
{\bf D}^{\alpha}_{\bf q_1} ({\bf V}_1  \rho_{1}). \]

The force ${\bf F}_{1}$ acts on the first particle.
It is a sum of the internal forces
$${\bf F}_{1k}=
{\bf F}({\bf q}_{1},{\bf p}_{1},{\bf q}_{k},{\bf p}_{k},t),$$ 
and the external force
${\bf F}^{e}_1={\bf F}^{e}({\bf q}_{1},{\bf p}_{1},t)$.
In the case of binary interaction, we have
\be \label{Fie2}
{\bf F}_{1}={\bf F}^{e}_1+\sum^{N}_{k=2} {\bf F}_{1k}. \ee
Using Eq. (\ref{Fie2}), the second term in Eq. (\ref{r1i2})
can be rewritten in the form
\[ \hat I[2,...,N] 
{\bf D}^{\alpha}_{\bf p_1} ({\bf F}_1  \rho_{N})   
=\hat I[2,...,N] \Bigl(
{\bf D}^{\alpha}_{\bf p_1} ({\bf F}^{e}_1  \rho_{N}) +
\sum^{N}_{k=2}
{\bf D}^{\alpha}_{\bf p_1} ({\bf F}_{1k}  \rho_{N} ) \Bigr) = \]
\be  \label{rr1i3} =
{\bf D}^{\alpha}_{\bf p_1} ({\bf F}^{e}_1  \rho_{1}) +
\sum^{N}_{k=2}
{\bf D}^{\alpha}_{\bf p_1} \hat I[2,...,N]
\Bigl( {\bf F}_{1k}  \rho_{N} \Bigr) . \ee
We assume that $\rho_N$ is invariant under the
permutations of identical particles \cite{Bog2}:
\[ \rho_N(...,{\bf q}_{k},{\bf p}_{k},...,{\bf q}_{l},{\bf p}_{l},...,t)=
\rho_N(...,{\bf q}_{l},{\bf p}_{l},...,{\bf q}_{k},{\bf p}_{k},...,t) . \]
In this case, $\rho_{N}$ is a symmetric function, 
and all $(N-1)$ terms of sum (\ref{rr1i3}) are identical.
Therefore the sum can be replaced by one term with the
multiplier $(N-1)$:
\be \label{rhs} \sum^{N}_{k=2}  \hat I[2,...,N] \
{\bf D}^{\alpha}_{\bf p_{1s}}
\Bigl( {\bf F}_{1k}  \rho_{N} \Bigr) 
= (N-1)  \hat I[2,...,N] \
{\bf D}^{\alpha}_{\bf p_1}
\Bigl( {\bf F}_{12}  \rho_{N} \Bigr)  . \ee
Using $\hat I[2,...,N]=\hat I[2]\hat I[3,...,N]$, 
we rewrite the right-hand side of (\ref{rhs}) in the form
\be \label{Eq77} \hat I[2] \
{\bf D}^{\alpha}_{\bf p_1} 
\Bigl( {\bf F}_{12} \hat I[3,...,N]  \rho_{N} \Bigr) 
={\bf D}^{\alpha}_{\bf p_1}
\hat I[2] \ \Bigl( {\bf F}_{12}  \rho_{2} \Bigr) , \ee
where $ \rho_2$ is two-particle density of probability
that is defined by the fractional integration
of the $N$-particle density of probability over all 
${\bf q}_{k}$ and ${\bf p}_{k}$, except $k=1,2$:
\be \label{p2}  \rho_{2}=
\rho({\bf q}_{1},{\bf p}_{1},{\bf q}_{2},{\bf p}_{2},t)=
\hat I[3,...,N]  \rho_{N}({\bf q},{\bf p},t) . \ee
Since ${\bf p}_{1}$ is independent of ${\bf q}_{2}$, ${\bf p}_{2}$, 
we can change the order of
the integrations and the differentiations:
\[ \hat I[2] \ {\bf D}^{\alpha}_{\bf p_1}
\Bigl( {\bf F}_{12}  \rho_{2} \Bigr) =
{\bf D}^{\alpha}_{\bf p_1}
\hat I[2] {\bf F}_{12}  \rho_{2}  . \]

Finally, we obtain 
\be \label{er1-2} \frac{\partial  \rho_{1}}{\partial t}+
{\bf D}^{\alpha}_{\bf q_1} ({\bf V}_1  \rho_{1})+
{\bf D}^{\alpha}_{\bf p_1} ({\bf F}^{e}_1  \rho_{1}) =
I( \rho_{2}). \ee
Here, $I( \rho_{2})$ is defined by 
\be \label{I2} I( \rho_{2})=
-(N-1) {\bf D}^{\alpha}_{\bf p_1}
\hat I[2] {\bf F}_{12}  \rho_{2} . \ee
The term $I( \rho_{2})$ describes a velocity of particle 
number change in $4m$-dimensional
two-particle elementary phase volume.
This change is caused by the interactions between particles. 

Equation (\ref{er1-2}) is the fractional generalization of the 
first Bogoliubov equation. 
If $\alpha=1$, then we have the first Bogoliubov equation for
non-Hamiltonian systems \cite{AP2005-1}.  
For Hamiltonian systems,
\be \label{Ham}
{\bf F}_{1}=-\frac{\partial H(q_1,p_1)}{\partial q_1}, \quad
{\bf V}_{1}=\frac{\partial H(q_1,p_1)}{\partial p_1}, \ee
and Eq. (\ref{er1-2}) has the well-known form \cite{Bog3,Bog3b,Gur,Petrina}.

\section{Vlasov equation with fractional derivatives}

Let us consider the particles as statistical independent systems
\cite{Vlasov,Vlasov2}. Then
\be \label{2-12}
 \rho_2({\bf q}_{1},{\bf p}_{1},{\bf q}_{2},{\bf p}_{2},t)=
 \rho_1({\bf q}_{1},{\bf p}_{1},t)
 \rho_1({\bf q}_{2},{\bf p}_{2},t) . \ee
Substitution of Eq. (\ref{2-12}) into Eq. (\ref{I2}) gives
\be \label{V-I} 
I( \rho_{2})=-{\bf D}^{\alpha}_{\bf p_1} \rho_1 
\hat I[2] {\bf F}_{12}  \rho_1({\bf q}_{2},{\bf p}_{2},t) , 
\ee
where $\rho_1=\rho_1({\bf q}_{1},{\bf p}_{1},t)$.
As a result, the effective force is
\[ {\bf F}^{eff} ({\bf q}_{1},{\bf p}_{1},t)=
\hat I[2] {\bf F}_{12}  \rho_{1}({\bf q}_{2},{\bf p}_{2},t). \]
In this case, we can rewrite Eq. (\ref{V-I}) in the form
\be \label{Ir2} I( \rho_{2})=
- {\bf D}^{\alpha}_{\bf p_1} ( \rho_{1} {\bf F}^{eff}) . \ee
Substituting of Eq. (\ref{Ir2}) into Eq. (\ref{er1-2}), we obtain
\be \label{p1-1} \frac{\partial  \rho_{1}}{\partial t}+
{\bf D}^{\alpha}_{\bf q_1} ({\bf V}_1  \rho_{1})
+{\bf D}^{\alpha}_{\bf p_1} \Bigl(
({\bf F}^{e}_1+(N-1){\bf F}^{eff}) \rho_{1} \Bigr)=0 . \ee
This equation is a closed equation for one-particle
density of probability with the external force ${\bf F}^{e}_1$
and the effective force ${\bf F}^{eff}$.
Equation (\ref{p1-1}) is the fractional generalization 
of the Vlasov equation that has coordinate 
derivatives of non-integer order.
For $\alpha=1$, we get the Vlasov equation for 
the non-Hamiltonian systems that is described 
by non-potential fields. 
For Hamiltonian systems (\ref{Ham}), equation 
(\ref{p1-1}) has the usual form \cite{Vlasov,Vlasov2}.

\section{Linear fractional kinetic equation 
for a system of charged particles}

Let us consider fractional kinetic equation (\ref{er1-2})
with $I(\rho_2)=0$, ${\bf V}_1={\bf p}/m={\bf v}$, and 
${\bf F}^e=e{\bf E}$, ${\bf B}=0$.
In this case, Eq. (\ref{er1-2}) has the form
\be \label{fke1} \frac{\partial f}{\partial t}+
({\bf v}, {\bf D}^{\alpha}_{\bf q} f) +
e({\bf E},  {\bf D}^{\alpha}_{\bf p} f) =0, \ee
where $f=\rho_1$ is one-particle density of probability, and
\be
({\bf v}, {\bf D}^{\alpha}_{\bf q} f) =
\sum^m_{s=1} (v_s, {\bf D}^{\alpha}_{{\bf q}_s} f) .
\ee

If we take into account the magnetic field (${\bf B}\not=0$), 
then we must use the fractional generalization of Leibnitz rules:
\be
{\bf D}^{\alpha}_{\bf p} (fg)=\sum^{\infty}_{r=0} 
\frac{\Gamma(\alpha+1)}{\Gamma(r+1) \Gamma(\alpha-r+1)}
({\bf D}^{\alpha-r}_{\bf p} f) D^r_{p} g ,
\ee
where $r$ are integer numbers. 
In this case, Eq. (\ref{fke1}) has the addition term 
\[ \frac{e}{mc}
{\bf D}^{\alpha}_{\bf p} \left( [{\bf p}, {\bf B}] f \right)=
\frac{e}{mc} \sum_{k l m} 
{\bf D}^{\alpha}_{p_k} \left( \varepsilon_{klm} p_l B_m f \right)
=\frac{e}{mc} \sum_{k l m} \varepsilon_{klm} B_m 
{\bf D}^{\alpha}_{p_k} \left( p_l f \right)= \]
\[ =\frac{e}{mc} \sum_{k l m} \varepsilon_{klm} B_m 
\sum^1_{r=0}\frac{\Gamma(\alpha+1)}{\Gamma(r+1) \Gamma(\alpha-r+1)}
[{\bf D}^{\alpha-i}_{p_k} f] \delta_{kl} p^r_l= \]

\[ =\frac{e}{mc} \sum_{k l m} \varepsilon_{klm} B_m  \left(
[{\bf D}^{\alpha}_{p_k} f] p_l+
\alpha [{\bf D}^{\alpha-1}_{p_k} f] \delta_{kl} \right)= \]

\be
 =\frac{e}{mc} \sum_{k l m} \varepsilon_{klm} B_m p_l
[{\bf D}^{\alpha}_{p_k} f ]=
\frac{e}{mc} \left( ({\bf D}^{\alpha}_{p_k} f),[{\bf p}, {\bf B}] \right)  .
\ee

Let us consider the perturbation \cite{Eck,KT} 
of the density of probability $f_0$ in the form
\be \label{pert}
f(t,q,p)=f_0 +\delta f(t,q,p) ,
\ee
where $f_0$ is a homogeneous stationary density of probability
that satisfies Eq. (\ref{fke1})  for ${\bf E}=0$.
Substituting of (\ref{pert}) into Eq. (\ref{fke1}), 
we get for the first perturbation 
\be \label{fke2} \frac{\partial \delta f}{\partial t}+
({\bf v}, {\bf D}^{\alpha}_{\bf q} \delta f) +
e({\bf E} , {\bf D}^{\alpha}_{\bf p} f_0 )=0. \ee
Equation (\ref{fke2}) is the linear fractional kinetic equation for 
the first perturbation of the density of probability.
Solutions of fractional linear kinetic equations of type (\ref{fke2})
are considered in Ref. \cite{SZ}.
For ${\bf E}=0$, the function $\delta f$ is described by 
\be \label{HH11}
(g_st)^{-1/\alpha} L_{\alpha} \left[ q_s (g_st)^{-1/\alpha} \right],
\ee 
where $g_s=v_s (D^{\alpha}_{q_s} q_s)^{-1}$, and
\be \label{La}
L_{\alpha}[x]=\frac{1}{2\pi} \int^{+\infty}_{-\infty} dk \  
e^{-ikx} e^{-|k|^{\alpha}} 
\ee
is the Levy stable density of probability \cite{Feller}.

For $\alpha=1$, the function (\ref{La}) gives
the Cauchy distribution 
\be  \label{HHH1}
L_1 [x]=\frac{1}{\pi} \frac{1}{x^2+1} ,
\ee
and Eq. (\ref{HH11}) is
\be \label{HHH2}
\frac{1}{\pi} \frac{ (g_s t)^{-1}}{q^2_s (g_st)^{-2} +1}  .
\ee 
For $\alpha=2$, we get the Gauss distribution: 
\be  \label{HHH3}
L_2 [x]= \frac{1}{2\sqrt{\pi}} e^{-x^2/4} ,
\ee
and the function (\ref{HH11}) is
\be \label{HHH4}
(g_st)^{-1/2} \frac{1}{2\sqrt{\pi}} e^{-q^2_s /(4g_s t) } .
\ee 

For $1< \alpha \le 2$, the function $L_{\alpha}[x]$ can be presented  
as the expansion
\be
L_{\alpha}[x]=-\frac{1}{\pi x} \sum^{\infty}_{n=1} 
(-x)^n \frac{\Gamma(1+n/\alpha)}{n!} \sin (n \pi/2) .
\ee
The asymptotic ($x \rightarrow \infty$, $1<\alpha<2$) is given by
\be
L_{\alpha}[x] \sim -\frac{1}{\pi x} \sum^{\infty}_{n=1} 
(-1)^n x^{-n \alpha} \frac{\Gamma(1+n \alpha)}{n!} \sin (n \pi/2)  .
\ee
As a result, we arrive at the asymptotic of the solution, 
which exhibits power-like tails for $x \rightarrow \infty$.
The tails is the important property of the solutions of
fractional equations.


\section{Conclusion}

In this paper, we consider equations with derivatives 
of non-integer order that can be used in statistical mechanics 
and kinetic theory.
We derive the Liouville, Bogoliubov and Vlasov 
equations with fractional derivatives 
with respect to coordinates and momenta. 
To derive the fractional Liouville equation, we consider the conservation
of probability to find a system in the fractional differential volume element.
Using the fractional Liouville equation, we obtain
the fractional generalization of the Bogoliubov hierarchy equations.
Fractional Bogoliubov equations can be used to derive 
fractional kinetic equations \cite{Zaslavsky1,Zaslavsky7,SZ}.
The fractional kinetic is related to equations 
that have derivatives of non-integer orders.
Fractional equations appear in the description 
of chaotic dynamics and fractal media.
For the fractional linear oscillator, the physical meaning
of the derivative of order $\alpha<2$ is dissipation.
Note that fractional derivatives with respect to coordinates
can be connected with long-range power-law interaction of 
the systems \cite{Lask,TZ3,KZT}.


\end{document}